# Electronic band structure, Fermi surface, and elastic properties of polymorphs of the new 5.2K iron-free superconductor SrPt$_2$As$_2$ from first principles calculations

I. R. Shein * and A. L. Ivanovskii

*Institute of Solid State Chemistry, Ural Branch, Russian Academy of Sciences, 91, Pervomaiskaya St., Ekaterinburg, 620990 Russia*

ABSTRACT

By means of the first-principles calculations, we studied in detail the structural, elastic, and electronic properties of the new tetragonal CaBe$_2$Ge$_2$-type 5.2K superconductor SrPt$_2$As$_2$ in comparison with two hypothetical SrPt$_2$As$_2$ polymorphs with ThCr$_2$Si$_2$-type structures, which differ in the atomic configurations of [Pt$_2$As$_2$] (or [As$_2$Pt$_2$]) blocks. We found that CaBe$_2$Ge$_2$-type SrPt$_2$As$_2$ is a unique system with near-Fermi bands of a complicated character and an "intermediate"-type Fermi surface, which consists of electronic pockets having a cylinder-like (2D) topology (typical of 122 FeAs phases) together with 3D-like electronic and hole pockets, which are characteristic of ThCr$_2$Si$_2$-like iron-free low-$T_c$ superconductors. Our analysis revealed that as distinct from ThCr$_2$Si$_2$-like 122 phases, other features of CaBe$_2$Ge$_2$-like SrPt$_2$As$_2$ are as follows: (1). essential differences in contributions from [Pt$_2$As$_2$] and [As$_2$Pt$_2$] blocks to the near-Fermi region when conduction is anisotropic and occurs mainly in [Pt$_2$As$_2$] blocks; (2). formation of a 3D system of strong covalent Pt-As bonds (inside and between [Pt$_2$As$_2$]/[As$_2$Pt$_2$] blocks), which is responsible for enhanced stability of this polymorph; and (3). essential charge anisotropy between adjacent [Pt$_2$As$_2$] and [As$_2$Pt$_2$] blocks. We also predicted that CaBe$_2$Ge$_2$-like SrPt$_2$As$_2$ is a mechanically stable and relatively soft material with high compressibility, which will behave in a ductile manner. On the contrary, the ThCr$_2$Si$_2$-type SrPt$_2$As$_2$ polymorphs, which contain only [Pt$_2$As$_2$] or [As$_2$Pt$_2$] blocks, are less stable, have Fermi surfaces of a multi-sheet three-dimensional type like the ThCr$_2$Si$_2$-like iron-free 122 phases, and therefore will be ductile materials with high elastic anisotropy. Based on our data for the three simplest SrPt$_2$As$_2$ polymorphs we assumed that there may exist a family of higher-order polytypes, which can be formed as a result of various stacking of the two main types of building blocks ([Pt$_2$As$_2$] and [As$_2$Pt$_2$]) in various combinations along the $z$ axis. This may provide an interesting platform for further theoretical and experimental search for new superconducting materials.

*PACS* : 71.18.+y, 71.15.Mb, 74.25.Jb

*Keywords:* Tetragonal SrPt$_2$As$_2$ polymorphs, Structural, electronic, elastic properties; *ab initio* calculations

# I. INTRODUCTION

Among the broad family of recently discovered [1] iron-based superconductors (SCs), the so-called 122 phases [2-9] belong to one of the most interesting and intensely studied groups of such materials. Their parent phases are $A$Fe$_2$As$_2$ (where $A$ are alkali earth metals Ca, Sr, and Ba or Eu), which adopt a quasi-two-dimensional (2D) tetragonal crystal structure of the ThCr$_2$Si$_2$-type, where [Fe$_2$As$_2$] blocks are separated by $A$ atomic sheets. In turn, inside [Fe$_2$As$_2$] blocks, Fe ions form a square lattice sandwiched between two As sheets shifted so that each Fe is surrounded by a distorted As tetrahedron {FeAs$_4$}.

The most remarkable aspects of 122 SCs are as follows: (i) undoped $A$Fe$_2$As$_2$ phases exhibit a collinear antiferromagnetic (AF) spin density wave (SDW), [2,5,10] (ii) superconductivity emerges owing to their hole or electron doping (or external pressure), [2-9] and (iii) near-Fermi Fe 3$d$-like bands play an essential role in superconductivity. [6,7,9]

Thus, the atomic substitutions inside [Fe$_2$As$_2$] blocks exert a profound influence on the properties of these 122 phases, in particular, on their superconductivity. So, it was found that electron doping of $A$Fe$_2$As$_2$ phases as a result of partial substitution $M \rightarrow$ Fe, where $M$ are some magnetic or non-magnetic 3$d$-5$d$ metals of VIII group (Co, Ni, Pd, Ru, Rh, or Ir), induces superconductivity with $T_c$ to ~ 20 K for such doped $A$Fe$_{2-x}M_x$As$_2$ phases.

The phase diagrams for such $A$Fe$_{2-x}M_x$As$_2$ systems usually show a SDW that competes with superconductivity; the onset of superconductivity coincides with a Lifshitz transition [10,11] (change in the Fermi surface (FS) topology) and superconductivity generally appears [12-16] in a broad window of $M$/Fe stoichiometry when the SDW is suppressed by doping. Besides, to gain a further insight into the nature of these systems, attention was paid also to the so-called "over-doped" limit, and a series of iron-free 122 phases $AM_2$As$_2$ ($M$ = Co, Ni, Pd, Ru, Rh, or Ir) has been recently examined both experimentally and theoretically. [17-27] All these $AM_2$As$_2$ phases (such as SrNi$_2$As$_2$, BaNi$_2$As$_2$, SrRu$_2$As$_2$, BaRu$_2$As$_2$, SrRh$_2$As$_2$, BaRh$_2$As$_2$ $etc.$) preserve the ThCr$_2$Si$_2$-like type, but show very low transition temperatures ranging in the interval $T_c$ ~ 0.3 - 3.0K.

In this context, of interest is the unusual situation for the systems $A$-(Fe/Pt)-As. Indeed, the diagrams of superconducting and magnetic phases in the Pt-substituted series BaFe$_{2-x}$Pt$_x$As$_2$ [28,29] and SrFe$_{2-x}$Pt$_x$As$_2$ [30] (for which the maximal $T_c$ to ~ 23K and ~ 16K was achieved at x = 0.05 [29] and x = 0.16, [30] respectively) are similar to those for the above transition-metal-substituted $A$Fe$_2$As$_2$ systems. These Pt-doped systems also keep the ThCr$_2$Si$_2$-like structure of initial $A$Fe$_2$As$_2$ phases.

On the contrary, the iron-free SrPt$_2$As$_2$, which may be viewed as the fully Pt-substituted $A$Fe$_2$As$_2$ as compared with other known 122-like iron-free SCs, shows the maximal $T_c$ ~ 5.2K [31] and adopts [31,32] a tetragonal CaBe$_2$Ge$_2$-type structure. Thus, for SrPt$_2$As$_2$ we observed a very intriguing structural situation. Indeed, the structures for the above ThCr$_2$Si$_2$-like $AM_2$As$_2$ may be described as a sequence of atomic sheets (along the $c$ axis) …–[As–$M_2$–As]–$A$–[As–$M_2$–As]–$A$–… On the contrary, for CaBe$_2$Ge$_2$-like SrPt$_2$As$_2$ this sequence is: …–[As–Pt$_2$–As]–Sr–[Pt–As$_2$–Pt]–Sr–[As–Pt$_2$–As]–Sr–…, $i.e.$ this structure contains [Pt$_2$As$_2$] blocks where each Pt atom is surrounded by an As tetrahedron {PtAs$_4$} alternating with [As$_2$Pt$_2$] blocks consisting of {AsPt$_4$} tetrahedrons. Thus, the new 5.2K iron-free superconductor SrPt$_2$As$_2$ cannot be considered as an analogue of the above ThCr$_2$Si$_2$-like iron-free 122 phases $AM_2$As$_2$ ($M$ = Co, Ni, Pd, Ru, Rh, or Ir).

In view of these circumstances, in the present work we performed the first-principles calculations to analyze the effects of the above unique structural features on the electronic and elastic properties for $SrPt_2As_2$.

The aim of this work was of two kinds. Firstly, we focused our attention on the peculiarities of electronic band structure, Fermi surface topology, and elastic properties for the recently synthesized [31,32] 5.2K SC $SrPt_2As_2$ with a tetragonal $CaBe_2Ge_2$-type structure. Secondly, using $SrPt_2As_2$ as an example, we wanted to determine how the stability and some properties of 122-like phases are affected by local atomic arrangements inside [$M$-As] blocks, *i.e.* {$M$As$_4$} *versus* {As$M_4$}, and by various stacking of these blocks, when external atoms from neighboring blocks can form different types of inter-blocks bonds, namely, As-$M$, As-As or $M$-$M$.

For this purpose, three tetragonal $SrPt_2As_2$ polymorphs: the synthesized $CaBe_2Ge_2$-like phase and two hypothetical $ThCr_2Si_2$-like phases with alternative atomic configurations of [$Pt_2As_2$] (or [$As_2Pt_2$]) blocks, have been examined. As a result, the structural parameters, stability, electronic bands, FS topology, densities of states, and the peculiarities of inter-atomic interactions for the above polymorphs have been obtained and analyzed. In addition, the elastic parameters (independent elastic constants, bulk and shear moduli, indexes of elastic anisotropy) have been predicted for these polymorphs, and the Young's moduli, Poisson's ratio, and Pugh's indicator of brittle/ductile behavior for the corresponding polycrystalline systems (in the Voigt-Reuss-Hill approximation) have been evaluated.

## II. MODELS AND COMPUTATIONAL ASPECTS

The synthesized [31,32] $SrPt_2As_2$ crystallizes in a tetragonal $CaBe_2Ge_2$-type structure (space group P4/nmm, #129). The atomic positions are Sr: 2c (¼, ¼, $z_{Sr}$); 2a (¾, ¼, 0); Pt2: 2c (¼, ¼, $z_{Pt}$); As1: 2b (¾, ¼, ½); and As2: 2c (¼, ¼, $z_{As}$), where $z_{Sr,Pt,As}$ are the so-called internal coordinates. The structure of this phase (further denoted as SPA-I) can be schematically described as a sequence of Sr sheets and [$Pt_2As_2$] and [$As_2Pt_2$] blocks consisting of {$PtAs_4$} and {$AsPt_4$} tetrahedrons: …[$Pt_2As_2$]/Sr/[$As_2Pt_2$]/Sr/[$Pt_2As_2$]/Sr/[$As_2Pt_2$]… as shown in Fig. 1.

We also examined two hypothetical tetragonal $SrPt_2As_2$ polymorphs with a $ThCr_2Si_2$-type structure (space group *I*4/*mmm*; #139). For one of them, denoted as SPA-II, the "conventional" $ThCr_2Si_2$-type structure was chosen, where the atomic positions are Sr: 2a (0, 0, 0), Pt: 4d (½, 0, ½) and As: 4e (0, 0, $z_{As}$); here, the stacking sequence is …Sr/[$Pt_2As_2$]/Sr/[$Pt_2As_2$]/Sr… For the second hypothetical polymorph (abbreviated as SPA-III) we also used the $ThCr_2Si_2$ structural type, but with inverse distribution of Pt and As over the atomic sites (in blocks) as compared with SPA-II, *i.e.* the atomic positions here are: Sr: 2a (0, 0, 0), As: 4d (½, 0, ½) and Pt: 4e (0, 0, $z_{Pt}$). The stacking sequence for SPA-III is … Sr/[$As_2Pt_2$]/[$As_2Pt_2$]/Sr…, see Fig. 1.

As a result, three basic tetragonal types of $SrPt_2As_2$ polymorphs have been examined, which enable us to clarify the role of (i) local atomic arrangement inside [Pt-As] blocks, *i.e.* {$PtAs_4$} *versus* {$AsPt_4$}, and of (ii) the main types of stacking of these blocks, *i.e.* …[$Pt_2As_2$]/Sr/[$As_2Pt_2$]/Sr/[$Pt_2As_2$]/Sr/[$As_2Pt_2$]… *versus* …Sr/[$Pt_2As_2$]/Sr/[$Pt_2As_2$]/Sr… *versus* …Sr/[$As_2Pt_2$]/[$As_2Pt_2$]/Sr…, when the external atoms from neighboring blocks can form various inter-blocks bonds, namely, As-Pt, As-As, and Pt-Pt, respectively.

Our band-structure calculations were carried out by means of the full-potential method with mixed basis APW+lo (LAPW) implemented in the WIEN2k suite of programs.[33] The generalized gradient correction (GGA) to exchange-correlation potential in the PBE form [34] was used. The plane-wave expansion was taken to $R_{MT} \times K_{MAX}$ equal to 8, and the $k$ sampling with 14×14×14 $k$-points in the Brillouin zone was used. The MT sphere radii were chosen to be 2.3 a.u. for Pt, 2.5 a.u. for Sr, and 1.9 a.u. for As. The calculations were performed with full-lattice optimization including internal coordinates. The self-consistent calculations were considered to be converged when the difference in the total energy of the crystal did not exceed 0.1 mRy and the difference in the total electronic charge did not exceed 0.001 $e$ as calculated at consecutive steps. On the example of SPA-I we have examined the influence of relativistic effects (the spin-orbital interactions (SOC) within FLAPW) on the valence bands and FS. It was found that SOC mainly results in energy shift and splitting of core and semi-core Pt states, which lay deeply under Fermi's level, whereas the common picture of valence bands (as well as the DOSs distributions and FS topology) as obtained in our calculations without SOC and within spin-orbit coupling varies very little. Thus further in all our calculations a standard procedure of electronic structure calculation and elastic properties in a scalar-relativistic approximation was used.

The hybridization effects were analyzed using the densities of states (DOSs), which were obtained by the modified tetrahedron method.[35] The ionic bonding was considered using Bader [36] analysis. In this approach, each atom of a crystal is surrounded by an effective surface that runs through minima of the charge density, and the total charge of an atom (the so-called Bader charge, $Q^B$) is determined by integration within this region. In addition, some peculiarities of intra-atomic bonding picture were visualized by means of charge density maps.

Furthermore, for the calculations of the elastic parameters of the considered SrPt$_2$As$_2$ polymorphs we employed the Vienna *ab initio* simulation package (VASP) in projector augmented waves (PAW) formalism.[37,38] Exchange and correlation were described by a nonlocal correction for LDA in the form of GGA.[34] The kinetic energy cutoff of 500 eV and k-mesh of 14×14×7 were used. The geometry optimization was performed with the force cutoff of 1 meV/Å.

These two DFT-based codes are complementary and allowed us to perform a complete investigation of the declared properties of the above systems.

### III. RESULTS AND DISCUSSION

#### A. Structural properties and stability

As the first step, the total energy ($E_{tot}$) *versus* cell volume calculations were carried out to determine the equilibrium structural parameters for the considered SrPt$_2$As$_2$ polymorphs. The calculated values are presented in Table I and are in reasonable agreement with the available experiments.[31,32] Some divergences are related to the well-known overestimation of the lattice parameters within LDA-GGA based calculation methods.

For the CaBe$_2$Ge$_2$-type polymorph (SPA-I), the Pt-As bond lengths (2.52 Å) between neighboring [Pt$_2$As$_2$]/[As$_2$Pt$_2$] blocks are comparable with those for the Pt-As lengths (about 2.6 Å) inside these blocks. Thus, this polymorph may be viewed as a quite isotropic phase with a three-dimensional (3D) system of strong Pt-As bonds. On the contrary, for the ThCr$_2$Si$_2$-type polymorphs (SPA-II and SPA-III, where the external planes are formed by As (or Pt) atoms within [As$_2$Pt$_2$] (or [Pt$_2$As$_2$]) blocks, respectively) the nearest As-As (or Pt-Pt) distances between neighboring blocks are twice smaller than the corresponding inter-atomic distances inside the blocks. On the other hand, for both polymorphs the Pt-As lengths inside [As$_2$Pt$_2$] (or [Pt$_2$As$_2$]) blocks are comparable with those for SPA-I, Table I. Thus, these simple crystallographic reasons allow us to expect that in contrast to SPA-I with a 3D system of Pt-As bonds, the bonding for SPA-II and SPA-III should be very anisotropic: simultaneously with Pt-As bonds inside the blocks, direct As-As (for SPA-II) or Pt-Pt bonds (for SPA-III) will be formed between the corresponding blocks, see also below.

Let us note also that the calculated bond angles Pt-As-Pt and As-Pt-As in tetrahedrons {PtAs$_4$} and {AsPt$_4$} for all of the SrPt$_2$As$_2$ polymorphs, see Table II, are far from the ideal tetrahedron angle (109.5°), which is considered as a factor favorable for superconductivity in FeAs systems. [6-9]

Next, the total-energy differences (ΔE) between the examined SrPt$_2$As$_2$ polymorphs are summarized in Table I. The results reveal that both within the FLAPW and VASP approaches the most stable and unstable polymorphs are, respectively, SPA-I with the CaBe$_2$Ge$_2$-like structure and SPA-II with a "conventional" ThCr$_2$Si$_2$-type lattice.

## B. Electronic band structure and Fermi surface

The calculated band structure and electronic densities of states (DOS) for the considered SrPt$_2$As$_2$ polymorphs are shown in Figs. 2 and 3, respectively. For all the polymorphs, their electronic spectra show some common features, namely (i) the As 4$p$ states occur between -7 eV and -4 eV with respect to the Fermi energy ($E_F$= 0 eV); (ii) most of the bands between -4 eV and $E_F$ are mainly of the Pt 5$d$ character, and (iii) the contributions from the valence states of Sr to the occupied bands are quite small. However, in the vicinity of the Fermi energy the topology of the electronic bands for various SrPt$_2$As$_2$ polymorphs becomes completely different.

So, for the synthesized CaBe$_2$Ge$_2$-like phase these near-Fermi bands demonstrate (Fig. 2) a unique complicated "mixed" character: simultaneously with quasi-flat bands along R-X, a series of high-dispersive Pt 5$d$ - like bands intersects the Fermi level between Γ and Z points and in the A-Z direction. These features yield an unusual multi-sheet FS, Fig. 4. Indeed, the Fermi surface of CaBe$_2$Ge$_2$-like SrPt$_2$As$_2$ consists of a set of hole and electronic sheets, where two electron-like pockets at the corners (around M) have a cylinder-like (2D) topology, and are very similar to the related electron cylinders along the $k_z$ direction at the zone corners obtained for ThCr$_2$Si$_2$-like $A$Fe$_2$As$_2$ materials. [6,7,39-42] However, the two other sheets (around Γ – electronic and hole-like sheets, see Fig. 5) are of a three-dimensional type similar to those for ThCr$_2$Si$_2$-like iron-free low-$T_c$ SCs such as SrNi$_2$As$_2$, SrRu$_2$As$_2$, BaRu$_2$As$_2$, SrRh$_2$As$_2$, *etc*. [23,24,43-45]

On the contrary, the Fermi surfaces of both ThCr$_2$Si$_2$-like SrPt$_2$As$_2$ polymorphs differ essentially from those of the FeAs-based 122 materials and are of a characteristic multi-sheet three-dimensional type like the above ThCr$_2$Si$_2$-like iron-free $AM_2$As$_2$ phases, [23,24,43-45] see Fig. 4.

As electrons near the Fermi surface are involved in the formation of the superconducting state, it is important to understand their nature. The total, atomic, and orbital decomposed partial DOSs at the Fermi level, N($E_F$), are shown in Table III. It is seen that for all SrPt$_2$As$_2$ polymorphs the main contribution to N($E_F$) comes from the Pt 5d states, with some additions of the As 4p states. For the examined polymorphs, the values of $N(E_F)$ decrease in the sequence: SPA-I > SPA-II > SPA-III, whereas the contributions of (As 4p)/(Pt 5d) states to $N(E_F)$ are: 0.303 (SPA-I) ~ 0.301 (SPA-II) < 0.370 (SPA-III).

The obtained data also allowed us to estimate the Sommerfeld constants ($\gamma$) and the Pauli paramagnetic susceptibility ($\chi$) for SrPt$_2$As$_2$ polymorphs under the assumption of the free electron model as $\gamma = (\pi^2/3)N(E_F)k_B^2$ and $\chi = \mu_B^2 N(E_F)$, Table III.

Note also that for the CaBe$_2$Ge$_2$-type SrPt$_2$As$_2$ polymorph (SPA-I) the contributions to $N(E_F)$ from the states of various blocks ([Pt$_2$As$_2$] *versus* [As$_2$Pt$_2$]) differ appreciably. Though the contributions of As1 4p and As2 4p states to $N(E_F)$ are quite small and comparable, the value of $N^{Pt1d}(E_F)$ = 0.59 states/eV·atom for Pt1 atoms (placed inside [Pt$_2$As$_2$] blocks) is twice greater than $N^{Pt2d}(E_F)$ = 0.30 states/eV·atom for Pt2 atoms located on the outer sides of [As$_2$Pt$_2$] blocks, see Fig. 3. Therefore, the conduction in SPA-I is expected to be most anisotropic, *i.e.* happening mainly in the [Pt$_2$As$_2$] blocks. To confirm this statement, we have calculated (within the FLMTO approach [46]) Fermi velocity of the carriers along different directions, $\langle v_{x,y,z}^2 \rangle^{1/2}$, for SrPt$_2$As$_2$ polymorphs, which together with the values of $\langle v_{x,y,z}^2 \rangle^{1/2}$ for some others layered SCs are presented in Table IV. We see that the ratio $\langle v_{x,y}^2 \rangle^{1/2} / \langle v_z^2 \rangle^{1/2}$ is maximal for SPA-I (1.52) - in comparison with SPA-II (1.03) and SPA-III (1.08), testifying the largest anisotropy of this system among examined polymorphs.

For 5.2K SC SrPt$_2$As$_2$, the experimental value of $\gamma^{exp}$ = 9.72 mJ·K$^{-2}$·mol$^{-1}$ has been evaluated [31] from standard analysis of specific heat $C(T)$ measurements. This allowed us to estimate the average electron-phonon coupling constant $\lambda$ for SrPt$_2$As$_2$ as $\gamma^{exp} = \gamma^{theor}(1 + \lambda)$. Within this crude estimation, the calculations yield $\lambda$ ~ 0.62, *i.e.* SrPt$_2$As$_2$ may be classified as a conventional phonon-mediated superconductor within a moderate coupling limit. For comparison, the available estimations of $\lambda$ for other related iron-free low-temperature 1111 and 122 SCs are about 0.58 (for LaNiPO [49]) or 0.76 (for BaNi$_2$As$_2$ [44]).

### C. Inter-atomic bonding

A conventional picture of inter-atomic interactions in ThCr$_2$Si$_2$-like $AM_2$As$_2$ phases assumes strong *M*-As bonding of a mixed ionic-covalent type inside [$M_2$As$_2$] blocks, some covalent As-As interactions between the adjacent [$M_2$As$_2$]/[$M_2$As$_2$] blocks together with ionic bonding between [$M_2$As$_2$] blocks and atomic $A$ sheets. [4,6-9,23,24,27,42,50,51] This bonding anisotropy determines the quasi-two-dimensional nature of these systems.

In our case, the overall character of ***covalent Pt-As bonding*** in SrPt$_2$As$_2$ polymorphs ***inside*** [Pt$_2$As$_2$] (or [As$_2$Pt$_2$]) blocks may be understood from site-projected DOS calculations. As is shown in Fig. 3, Pt 5d and As 4p states are strongly hybridized. In addition, the unoccupied levels (Pt 6p and As 3d) in the examined phases become partially occupied and will hybridized with others valence states, but their contributions in the valence band (and in covalent bonding) are quite small. On the other hand, a completely different bonding in the examined polymorphs arises ***between*** the adjacent blocks, and this situation is clearly visible in Fig. 6. So, while for CaBe$_2$Ge$_2$-like SrPt$_2$As$_2$ strong covalent Pt-As bonding takes place, for the hypothetical ThCr$_2$Si$_2$-like polymorphs the directed unipolar As-As (SPA-II) or Pt-Pt bonds (SPA-III) appear. Thus, it is possible to assume that this 3D system of strong covalent Pt-As bonds, which appear inside and between [Pt$_2$As$_2$]/[As$_2$Pt$_2$] blocks for CaBe$_2$Ge$_2$-like SrPt$_2$As$_2$, is responsible for enhanced stability of this polymorph.

In turn, for $AM_2As_2$ phases the ***ionic bonding*** is often explained [6-9,42,50] within the oversimplified ionic model. In our case, if we assume the usual oxidation numbers of atoms: $Sr^{2+}$, $Pt^{2+}$, and $As^{3-}$, the charge distributions for $SrPt_2As_2$ should be +2 for Sr sheets and -2 for $[Pt_2As_2]$ ($[As_2Pt_2]$) blocks. Thus, for all $SrPt_2As_2$ polymorphs: (i) both types of $[Pt_2As_2]$ or $[As_2Pt_2]$ blocks should adopt the same ionic states (2-), and (ii) the identical charge transfer (2*e*) occurs from $Sr^{2+}$ sheets to $[Pt_2As_2]^{2-}$ ($[As_2Pt_2]^{2-}$) blocks.

The real picture of charge distributions appears more complicated. To estimate numerically the amount of electrons redistributed between various atoms and between adjacent $[Pt_2As_2]^{n+}$ ($[As_2Pt_2]^{m+}$) blocks, we carried out a Bader [36] analysis. The total charge of an atom (the so-called Bader charge, $Q^B$), the corresponding charges as obtained from the purely ionic model ($Q^i$), and their differences ($\Delta Q = Q^B - Q^i$) are presented in Table V. The results confirm that all $SrPt_2As_2$ polymorphs are partly ionic compounds, and charges are transferred from Pt and Sr to As. Then, the effective charges for $[Pt_2As_2]$ ($[As_2Pt_2]$) blocks were found: $[Pt_2As_2]^{0.996-}$ *versus* $[As_2Pt_2]^{0.204-}$ for SPA-I, $[Pt_2As_2]^{0.578-}$ for SPA-II, and $[As_2Pt_2]^{0.576-}$ for SPA-III. Thus, in SPA-II and SPA-III we observed a quite identical charge transfer ($\delta \sim 0.6e$) from $Sr^{\delta+}$ sheets to $[Pt_2As_2]^{\delta-}$ ($[As_2Pt_2]^{\delta-}$) blocks. For SPA-I, on the contrary, essential charge anisotropy was obtained between adjacent $[Pt_2As_2]^{0.996-}/[As_2Pt_2]^{0.204-}$ blocks. Let us note that for $CaBe_2Ge_2$-type $SrPt_2As_2$ the distribution of non-equivalent ionic $[Pt_2As_2]^{0.996-}/[As_2Pt_2]^{0.204-}$ blocks around the $Sr^{\delta+}$ sheets is a quite rare situation, which takes place, for example, for another related low-$T_c$ SC – a layered phase $La_3Ni_4P_4O_2$ [52,53] with asymmetric distribution of ionic blocks around conducting $[Ni_2P_2]$ blocks.

### D. Elastic properties

Let us discuss the elastic parameters for the examined $SrPt_2As_2$ polymorphs as obtained within VASP calculations. The standard "volume-conserving" technique was used in the calculation of stress tensors on strains applied to the equilibrium structure to obtain the elastic constants $C_{ij}$. [54] In this way the values of six independent elastic constants for tetragonal crystals ($C_{11}$, $C_{12}$, $C_{13}$, $C_{33}$, $C_{44}$ and $C_{66}$) were estimated, Table VI.

First of all, $C_{ij}$ constants for all $SrPt_2As_2$ polymorphs are positive and satisfy the generalized criteria [55] for mechanically stable tetragonal materials: $C_{11} > 0$, $C_{33} > 0$, $C_{44} > 0$, $C_{66} > 0$, $(C_{11}-C_{12}) > 0$, $(C_{11}+ C_{33} - 2C_{13}) > 0$, and $[2(C_{11} + C_{12}) + C_{33} + 4C_{13}] > 0$. Note that for SPA-II the value of $C_{66}$ is very small. Thus, this polymorph (which is also energetically less favorable than SPA-I and SPA-III, according to total-energy calculations, Table I) lies on the border of mechanical stability.

Further, the calculated elastic constants $C_{ij}$ allowed us to obtain the bulk (*B*) and shear (*G*) moduli. Usually, for such calculations two main approximations are used, namely the Voigt (V) and Reuss (R) schemes, see for example Ref. [56]. So, in the Voigt approximation, these parameters are:
$$B_V = [2(C_{11} + C_{12}) + C_{33} + 4C_{13}]/9,$$
$$G_V = [M + 3C_{11} - 3C_{12} + 12C_{44} + 6C_{66}]/30,$$
in Reuss approximation:
$$B_R = C^2/M,$$
$$G_R = 15\{(18B_V/C^2) + [6/(C_{11} - C_{12})] + (6/C_{44}) + (3/C_{66})\}^{-1},$$
where M = $C_{11} + C_{12} + 2C_{33} - 4C_{13}$, and $C^2 = (C_{11} + C_{12})C_{33} - 2C_{13}^2$.

We evaluated also the corresponding parameters for polycrystalline $SrPt_2As_2$ species, *i.e.* for materials in the form of aggregated mixtures of microcrystallites with random orientation. For this purpose we utilized the Voigt-Reuss-Hill (VRH) approximation. [57-59] In this approach, the actual effective moduli ($B_{VRH}$ and $G_{VRH}$) for polycrystals are approximated by the arithmetic mean of the two above mentioned limits - Voigt and Reuss and further allowed us to obtain the Young's moduli $Y$ and the Poisson's ratio $v$ as: $Y_{VRH} = 9 B_{VRH}/\{1 + (3B_{VRH}/G_{VRH})\}$, and $v = (3B_{VRH} - 2G_{VRH})/2(3B_{VRH} + G_{VRH})$.

The above elastic parameters are presented in Table VII and allow us to make the following conclusions:

(i). The bulk moduli of the $SrPt_2As_2$ polymorphs increase in the sequence: $B$(SPA-II) < $B$(SPA-III) < $B$(SPA-I). As the bulk modulus represents the resistance to volume change against external forces, this indicates that the highest average bond strength will be achieved for SPA-I. On the other hand, the obtained bulk modulus for SPA-I is quite small (~ 100 GPa), and therefore the recently discovered 5.2K SC $SrPt_2As_2$ should be classified as a relatively soft material with high compressibility ($\beta \sim 0.01$ GPa$^{-1}$). In addition, the Young's modulus of materials is defined as a ratio of linear stress and linear strain, which tells about their stiffness. The Young's modulus of $CaBe_2Ge_2$-type $SrPt_2As_2$ was found to be $Y \sim 72$ GPa; thus, this material will show a rather small stiffness.

(ii). For all $SrPt_2As_2$ polymorphs it was found that $B > G$; this implies that the parameter limiting the mechanical stability of these materials is the shear modulus $G$, which represents the resistance to shear deformation against external forces.

(iii). One of the most widely used malleability measures of materials is the Pugh's criterion ($G/B$ ratio). [60] As is known empirically, if $G/B < 0.5$, a material behaves in a ductile manner, and *vice versa*, if $G/B > 0.5$, a material demonstrates brittleness. In our case, according to this indicator (Table VII), $SrPt_2As_2$ polymorphs will behave as ductile materials.

(iv). Elastic anisotropy of crystals reflects a different bonding character in different directions and has an important implication since it correlates with the possibility to induce microcracks in materials. [61,62] We have estimated the elastic anisotropy for the examined materials using the so-called universal anisotropy index [63] defined as: $A^U = 5G_V/G_R + B_V/B_R - 6$. For isotropic crystals $A^U = 0$; the deviations of $A^U$ from zero define the extent of crystal anisotropy. In our case, the minimal anisotropy is exhibited by 3D-like SPA-I, while SPA-II and SPA-III demonstrate the maximal (and comparable) deviations from $A^U = 0$ (Table VII) that testify to their high elastic anisotropy.

Finally, let us note that the elastic parameters of the 3D-like SPA-I polymorph (*i.e.* the synthesized 5.2K SC $SrPt_2As_2$ with a tetragonal $CaBe_2Ge_2$-type structure) are higher than those for $ThCr_2Si_2$-type FeAs phases. So, according to the available experimental and theoretical data, the bulk moduli are $B \sim 62$ GPa, [64] for $SrFe_2As_2$ and $B \sim 60$ GPa [65] for $CaFe_2As_2$ *versus* $B \sim 100$ GPa for $CaBe_2Ge_2$-type $SrPt_2As_2$ as obtained by us within VASP calculations. However, as a whole the elastic properties of 5.2K SC $SrPt_2As_2$ are comparable with the same for some other related FeAs SC's. So, the bulk moduli for 111 and 1111 FeAs SCs vary in the interval from ~ 57 GPa (for LiFeAs [66]) to ~ 100÷120 GPa for some 1111 FeAs phases such as $LaFeAsO$ or $NdFeAsO$. [64-69]

## IV. CONCLUSIONS

In summary, by means of the first-principles calculations we studied in detail the structural, elastic, and electronic properties of the new low-temperature superconductor – tetragonal CaBe$_2$Ge$_2$-type SrPt$_2$As$_2$ – in comparison with two hypothetical SrPt$_2$As$_2$ polymorphs with ThCr$_2$Si$_2$-type structures, which differ in the atomic configurations of [Pt$_2$As$_2$] (or [As$_2$Pt$_2$]) blocks.

Our studies showed that CaBe$_2$Ge$_2$-type SrPt$_2$As$_2$ is a unique system with near-Fermi bands of a complicated "mixed" character: simultaneously with quasi-flat bands a set of high-dispersive bands intersects the Fermi level. The Fermi surface of this phase adopts an "intermediate" character and consists of electronic pockets having a cylinder-like (2D) topology (typical of 122 FeAs phases) together with 3D-like electronic and hole pockets, which are characteristic of ThCr$_2$Si$_2$-like iron-free low-$T_c$ SCs such as SrNi$_2$As$_2$, SrRu$_2$As$_2$, SrRh$_2$As$_2$, *etc*. Next, the main contribution to N(E$_F$) comes from the Pt 5$d$ states, with some additions of the As 4$p$ states, but these contributions from the states of [Pt$_2$As$_2$] blocks are almost twice greater than from the [As$_2$Pt$_2$] blocks. Thus, conduction in CaBe$_2$Ge$_2$-type SrPt$_2$As$_2$ is expected to be anisotropic and happening mainly in [Pt$_2$As$_2$] blocks.

Further, in contrast to ThCr$_2$Si$_2$-like 122 phases, another feature of CaBe$_2$Ge$_2$-like SrPt$_2$As$_2$ is the type of intra-atomic bonding. Here, (1). a 3D system of strong covalent Pt-As bonds (inside and between [Pt$_2$As$_2$]/[As$_2$Pt$_2$] blocks) appears, which is responsible for enhanced stability of this polymorph, and (2). essential charge anisotropy was obtained between the adjacent [Pt$_2$As$_2$] and [As$_2$Pt$_2$] blocks. Finally, our analysis shows that the synthesized CaBe$_2$Ge$_2$-like SrPt$_2$As$_2$ is a mechanically stable and relatively soft material with high compressibility ($\beta \sim 0.01$ GPa$^{-1}$), which will behave in a ductile manner. However, this system will show the minimal anisotropy and an enhanced bulk modulus as compared with the examined ThCr$_2$Si$_2$-like polymorphs and other isostructural 122 phases.

In turn, the considered hypothetical tetragonal SrPt$_2$As$_2$ polymorphs with a ThCr$_2$Si$_2$-type structure, which contain only [Pt$_2$As$_2$] or [As$_2$Pt$_2$] blocks, are less stable. This fact can be explained taking into account the anisotropy of intra- and inter-blocks bonding, where alongside with Pt-As intra-blocks bonds, a system of unipolar As-As (or Pt-Pt) inter-blocks bonds appears. The Fermi surfaces of both ThCr$_2$Si$_2$-like SrPt$_2$As$_2$ polymorphs differ essentially from those of the CaBe$_2$Ge$_2$-like phase and are of a multi-sheet three-dimensional type like the above ThCr$_2$Si$_2$-like iron-free *AM*$_2$As$_2$ phases. Finally, our analysis reveals that these ThCr$_2$Si$_2$-like polymorphs will be ductile materials with high elastic anisotropy.

In conclusion, our calculations indicate that the difference in the ground-state energies of the examined SrPt$_2$As$_2$ phases is rather small. This allows us to speculate about the existence of a family of higher-order polytypes, which can be formed (by keeping the tetragonal type of the lattice) only as a result of various stacking of two main types of building blocks ("direct" [Pt$_2$As$_2$] blocks formed from tetrahedrons {PtAs$_4$} and "inverse" [As$_2$Pt$_2$] blocks formed from tetrahedrons {AsPt$_4$}) in various combinations along the *z* axis. We also assume a possibility of polytypism for other related 122 phases. These systems can provide an interesting platform for further theoretical and experimental search for new superconducting materials. Another aspect, which calls for further studies, is the effect of structure [32] modulation on the electronic properties of SrPt$_2$As$_2$.

## ACKNOWLEDGMENTS

This work was supported by the Russian Foundation for Basic Research, Grants No. RFBR- 09-03-00946 and No. RFBR- 10-03-96008.

TABLE I. The optimized lattice parameters (*a* and *c*, in Å), internal coordinates ($z_{Sr,Pt,As}$), some inter-atomic distances (*d*, in Å), and total-energy differences (ΔE, eV/cell) for the examined SrPt$_2$As$_2$ polymorphs.

| phase/parameter * | SPA-I | SPA-II | SPA-III |
|---|---|---|---|
| *a* | 4.5279/4.5135 (4.46-4.51) ** | 4.5218/4.5440 | 4.6428/4.6761 |
| *c* | 10.0137/10.0549 (9.81) ** | 10.1957/10.1852 | 9.7329/9.6393 |
| *c/a* | 2.2116/2.2277 | 2.2548/2.2415 | 2.0963/2.0614 |
| $z_{Sr,As,Pt,}$ | $z_{Sr}$=0.2504/0.2501 $z_{Pt}$=0.3810/0.3798 $z_{As}$=0.1286/0.1295 | $z_{As}$=0.3675/0.3665 | $z_{Pt}$=0.3599/0.3564 |
| $d^1$ | 2.60/2.61 (Pt-As in [Pt$_2$As$_2$]) 2.56/2.56 (Pt-As in [As$_2$Pt$_2$]) | 2.56/2.57 (Pt-As in [Pt$_2$As$_2$]) 4.00/4.01 (As-As in [Pt$_2$As$_2$]) | 2.78/2.76 (Pt-As in [As$_2$Pt$_2$]) 3.91/3.89 (Pt-Pt in [As$_2$Pt$_2$]) |
| $d^2$ | 2.52/2.52 (Pt-As) | 2.70/2.70 (As-As) | 2.76/2.78 (Pt-Pt) |
| ΔE | 0 | 0.46/0.42 | 0.20/0.16 |

\* as obtained within FLAPW/VASP
\*\* available experimental data (Ref. [31]) are given in parentheses
$d^1$ are the labeled inter-atomic distances inside [Pt$_2$As$_2$] ([As$_2$Pt$_2$]) blocks
$d^2$ are the nearest Pt-As (As-As, or Pt-Pt) distances between neighboring blocks: [Pt$_2$As$_2$]/[As$_2$Pt$_2$] ([Pt$_2$As$_2$]/[Pt$_2$As$_2$] or [As$_2$Pt$_2$]/[As$_2$Pt$_2$])

TABLE II. Calculated bond angles in {PtAs$_4$} ({AsPt$_4$}) tetrahedrons for the examined SrPt$_2$As$_2$ polymorphs.

| phase/parameter * | | SPA-I | SPA-II | SPA-III |
|---|---|---|---|---|
| {PtAs$_4$} | As-Pt-As | 104.2/104.5 | 102.7/102.5 | |
| | | 120.7/120.0 | 124.2/124.5 | |
| | Pt-As-Pt | 75.8/75.5 | 77.3/77.5 | |
| | | 120.7/120.0 | 124.4/124.5 | |
| {AsPt$_4$} | As-Pt-As | 77.4/77.1 | | 79.7/80.7 |
| | | 124.4/123.6 | | 130.0/133.0 |
| | Pt-As-Pt | 102.6/102.9 | | 100.3/99.1 |
| | | 124.4/123.6 | | 130.0/133.0 |

\* as obtained within FLAPW/VASP

TABLE III. Total (in states/eV·f.u.) and partial (in states/eV·atom) densities of states at the Fermi level, electronic heat capacity γ (in mJ·K$^{-2}$·mol$^{-1}$), and molar Pauli paramagnetic susceptibility χ (in 10$^{-4}$ emu·mol$^{-1}$) for the examined SrPt$_2$As$_2$ polymorphs.

| phase/parameter | As 4$p$ | Pt 5$d$ | total | γ | χ |
|---|---|---|---|---|---|
| SPA-I | 0.11/0.16 * | 0.59/0.30 * | 2.55 | 6.01 | 0.82 |
| SPA-II | 0.22 | 0.73 | 2.06 | 4.86 | 0.66 |
| SPA-III | 0.17 | 0.46 | 1.69 | 3.98 | 0.54 |

* for non-equivalent atoms: (Pt$^1$,As$^1$)/(Pt$^2$,As$^2$); see Sec. II.

TABLE IV. Fermi velocity along different directions ($\langle v_{x,y,z}^2 \rangle^{1/2}$ ×10$^7$ cm/s) for the examined SrPt$_2$As$_2$ polymorphs in comparison with some others layered SCs.

| phase | $\langle v_x^2 \rangle^{1/2}$ | $\langle v_y^2 \rangle^{1/2}$ | $\langle v_z^2 \rangle^{1/2}$ |
|---|---|---|---|
| SPA-I | 2.47 | 2.47 | 1.62 |
| SPA-II | 1.93 | 1.93 | 1.87 |
| SPA-III | 2.12 | 2.12 | 1.95 |
| MgB$_2$ * | 4.90 | 4.90 | 4.76 |
| La$_{1.85}$Sr$_{0.15}$CuO$_4$ ** | 2.2 | 2.2 | 0.41 |
| YBa$_2$Cu$_3$O$_7$ ** | 1.8 | 2.8 | 0.7 |

* Ref. [47].
** Ref. [48].

TABLE V. Effective atomic charges (in $e$) for SrPt$_2$As$_2$ polymorphs as obtained from a purely ionic model (Q$^i$), Bader analysis (Q$^B$), and their differences (ΔQ = Q$^B$-Q$^i$).

| phase/parameter | Q | Pt | As | Sr |
|---|---|---|---|---|
| | Q$^i$ | +2 | -3 | +2 |
| SPA-I | Q$^B$ | 10.565/10.894 * | 4.938/5.004 * | 8.599 |
| | ΔQ | 2.565/2.894 * | -3.063/-2.996 * | 0.599 |
| | Q$^i$ | +2 | -3 | +2 |
| SPA-II | Q$^B$ | 10.684 | 5.023 | 8.587 |
| | ΔQ | 2.684 | -2.973 | 0.587 |
| | Q$^i$ | +2 | -3 | +2 |
| SPA-III | Q$^B$ | 10.760 | 4.952 | 8.577 |
| | ΔQ | 2.760 | -3.048 | 0.577 |

* for non-equivalent atoms: (Pt$^1$,As$^1$)/(Pt$^2$,As$^2$); see Sec. II.

TABLE VI. Calculated elastic constants ($C_{ij}$, in GPa) for SrPt$_2$As$_2$ tetragonal polymorphs.

| phase/parameter | SPA-I | SPA-II | SPA-III |
|---|---|---|---|
| $C_{11}$ * | 136 | 132 | 165 |
| $C_{12}$ | 66 | 22 | 57 |
| $C_{13}$ | 89 | 56 | 69 |
| $C_{33}$ | 148 | 109 | 139 |
| $C_{44}$ | 30 | 31 | 25 |
| $C_{66}$ | 17 | < 5 | 27 |

* as obtained within VASP

TABLE VII. Calculated elastic parameters for SrPt$_2$As$_2$ polymorphs: bulk moduli ($B$, in GPa), compressibility ($\beta$, in GPa$^{-1}$), shear moduli ($G$, in GPa), Pugh's indicator ($G/B$), Young's moduli ($Y$, in GPa), Poisson's ratio ($v$), and the so-called universal anisotropy index ($A^U$).

| Phase/parameter * | SPA-I | SPA-II | SPA-III |
|---|---|---|---|
| $B_V$ | 101 | 71 | 95 |
| $B_R$ | 99 | 71 | 95 |
| $B_{VRH}$ | 100 | 71 | 95 |
| $\beta$ | 0.010 | 0.014 | 0.010 |
| $G_V$ | 27 | 29 | 34 |
| $G_R$ | 25 | 5 | 30 |
| $G_{VRH}$ | 26 | 17 | 32 |
| $G/B$ | 0.26 | 0.24 | 0.34 |
| $Y$ | 72 | 46 | 87 |
| $v$ | 0.38 | 0.39 | 0.35 |
| $A^U$ | 0.42 | -5.83 | -5.12 |

* as obtained within VASP

**Figures**

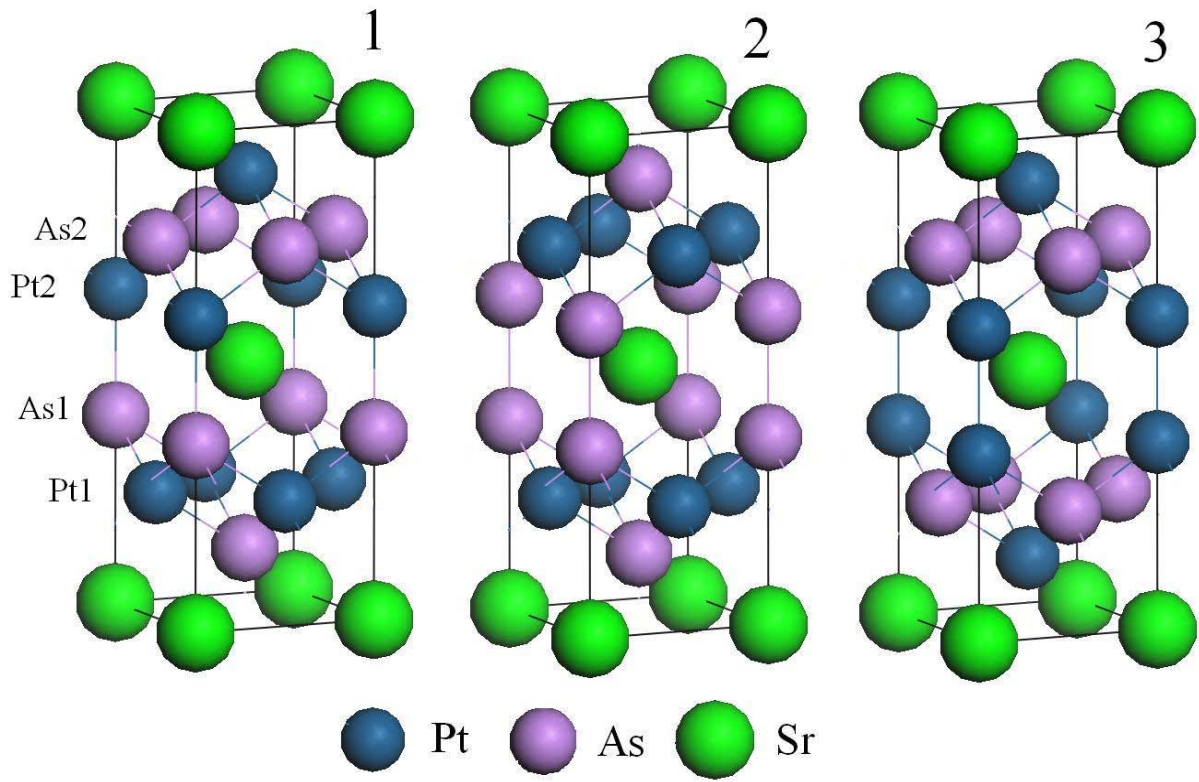

FIG. 1. (*Color online*) Crystal structures of the examined SrPt$_2$As$_2$ polymorphs: (1) SPA-I, (2) SPA-II, and (3) SPA-III (*see text*).

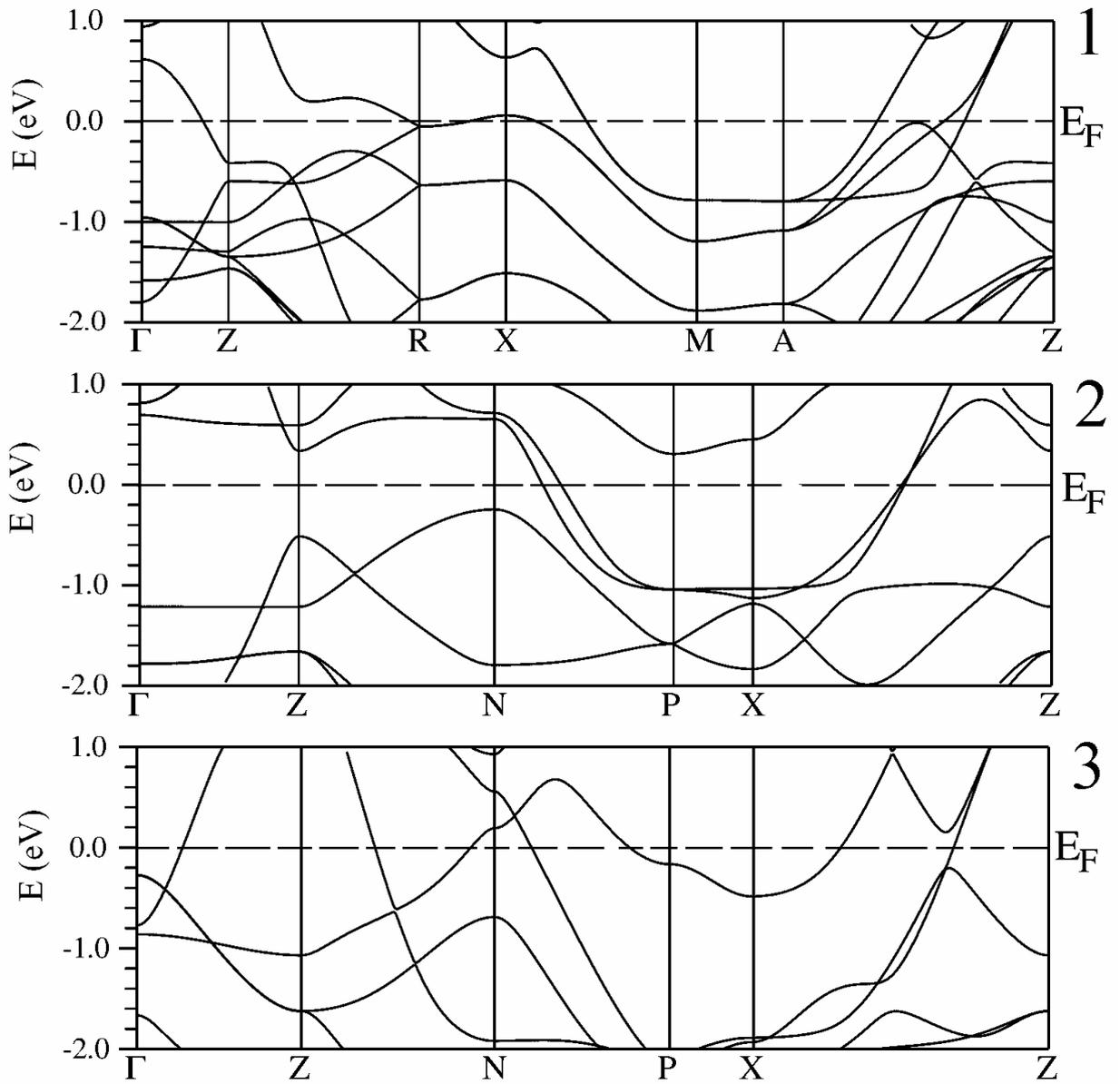

FIG. 2. Electronic band structures of SrPt$_2$As$_2$ polymorphs: (1) SPA-I, (2) SPA-II, and (3) SPA-III.

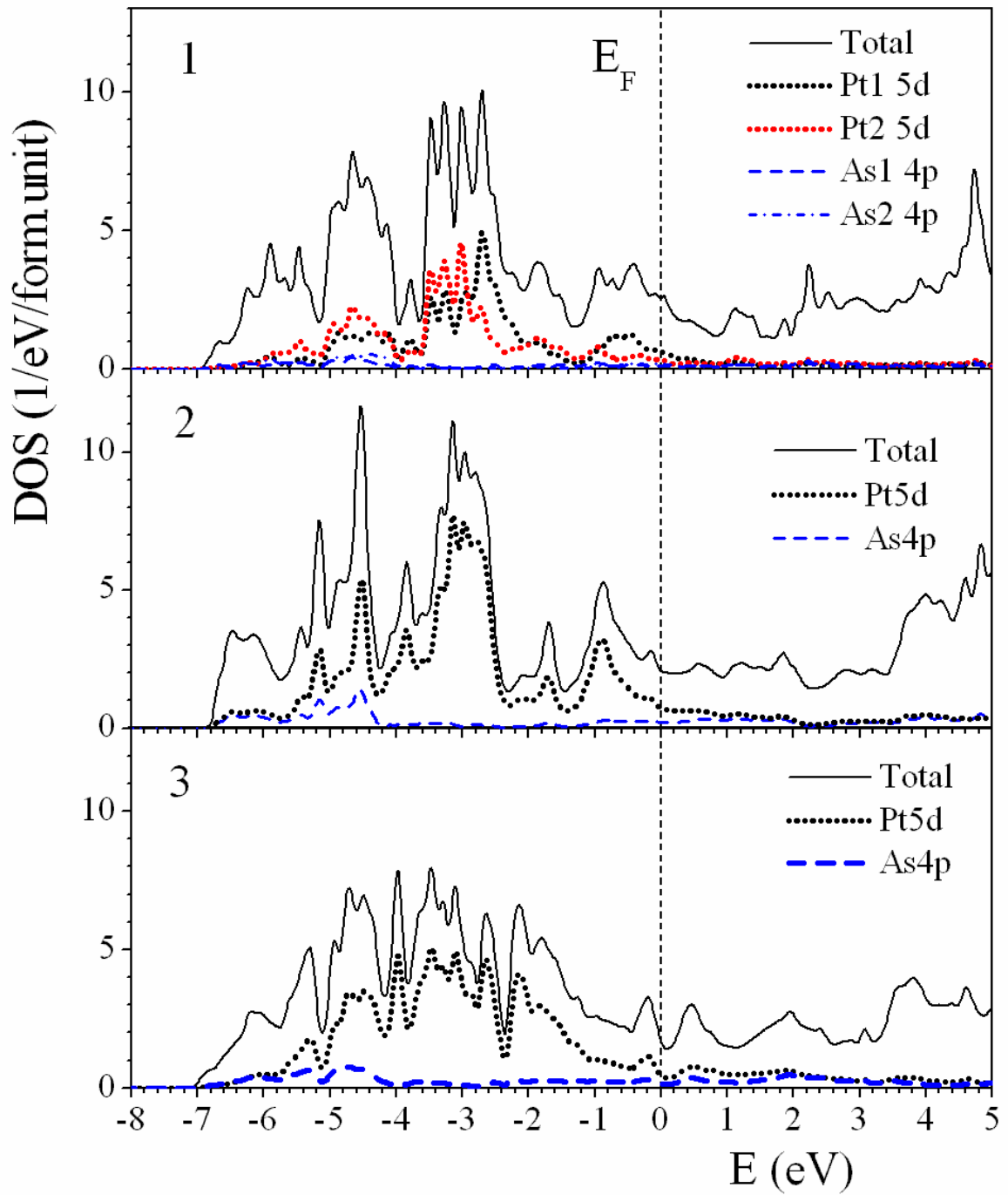

FIG. 3. (*Color online*) Total and partial densities of states of SrPt$_2$As$_2$ polymorphs: (1) SPA-I, (2) SPA-II, and (3) SPA-III.

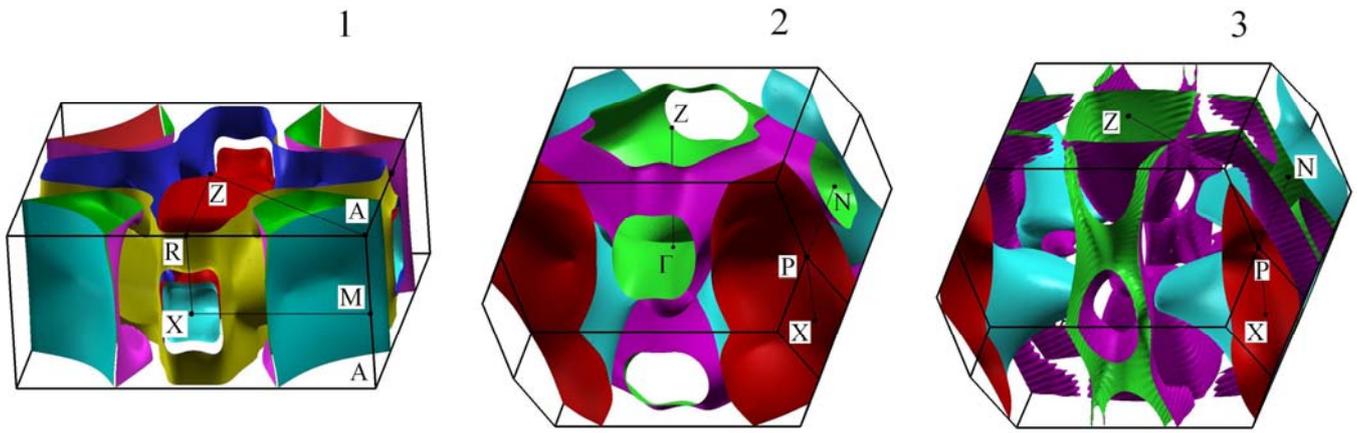

FIG. 4. (*Color online*) The Fermi surfaces of SrPt$_2$As$_2$ polymorphs: (1) SPA-I, (2) SPA-II, and (3) SPA-III.

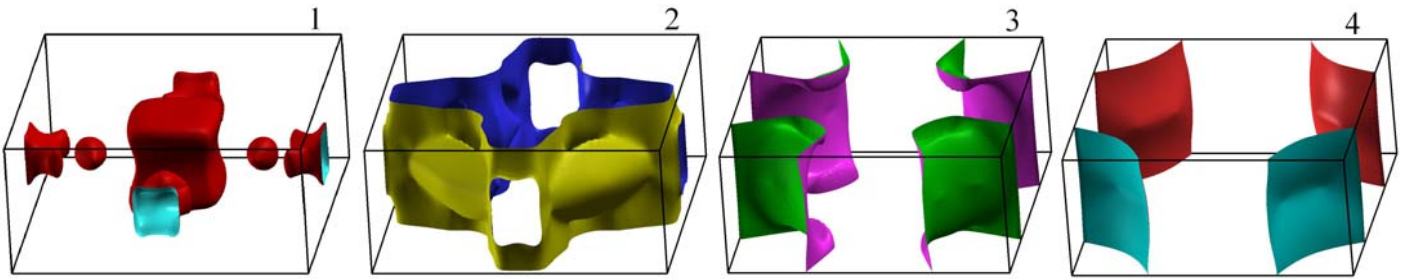

FIG. 5. (*Color online*) Separate sheets of the Fermi surface for CaBe$_2$Ge$_2$-like SrPt$_2$As$_2$. Three of them (2, 3, and 4) are electronic-like, and sheet 1 is hole-like, see also Ref. [45].

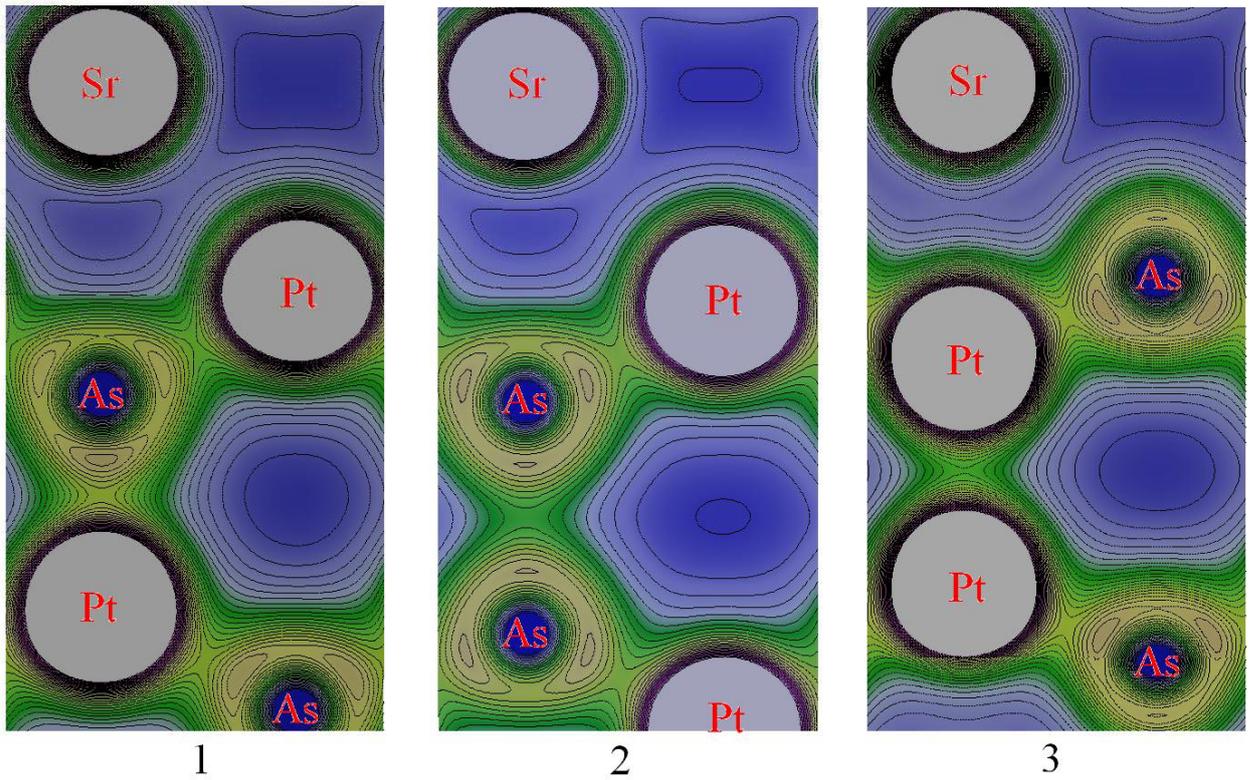

FIG. 6. (*Color online*) Charge density maps of SrPt$_2$As$_2$ polymorphs illustrating the formation of directional "inter-blocks" covalent bonds: (1) As-Pt bonds for SPA-I, (2) As-As bonds for SPA-II, and (3) Pt-Pt bonds –for SPA-III.